\documentstyle[preprint,aps]{revtex}
\tightenlines
\begin{document}
\draft
\title{A quantum-geometrical description of fracton statistics}

\author{Wellington da Cruz}
\address{Departamento de F\'{\i}sica,\\
 Universidade Estadual de Londrina, Caixa Postal 6001,\\
Cep 86051-970 Londrina, PR, Brazil\\
E-mail address: wdacruz@exatas.uel.br}
\date{\today}
\maketitle
\begin{abstract}

We consider the fractal characteristic of the quantum mechanical paths 
and we obtain for any universal class of fractons labeled by 
the Hausdorff dimension defined within the interval 
$1$$\;$$ < $$\;$$h$$\;$$ <$$\;$$ 2$, a fractal distribution function 
associated with a fractal von Neumann entropy. Fractons are 
charge-flux systems defined in two-dimensional multiply connected space 
and they carry rational or irrational values of spin. 
This formulation can be considered in the context of the fractional 
quantum Hall effect-FQHE and number theory.
\\

keywords: Fractal distribution function; fractal von Neumann entropy; 
fractional quantum Hall effect.
\\

Talk given at the 2nd 
International Londrina Winter 
School: Mathematical Methods in Physics, August, 26-30 (2002), Universidade 
Estadual de Londrina, Paran\'a, Brazil.
\end{abstract}
\newpage

\section{Introduction}	

We make out a review of some concepts introduced by us in the literature, 
such as\cite{R1,R2,R3,R4,R5,R6}: fractons, universal classes $h$ of particles, 
fractal spectrum, duality symmetry betwenn classes $h$ of particles, 
fractal supersymmetry, fractal distribution function, 
fractal von Neumann entropy, fractal index etc. We apply these 
ideas in the context of the FQHE and number theory. 

Fractons are charge-flux systems which carry rational or irrational 
values of spin. These objects are defined in 
two-dimensional multiply connected space and are classified in 
universal classes $h$ of particles or quasiparticles, with the 
fractal parameter or Hausdorff dimension $h$ , defined in the interval 
$1$$\;$$ < $$\;$$h$$\;$$ <$$\;$$ 2$. It is related to the 
quantum paths and can be extracted from the propagators 
of the particles in the momentum space\cite{R7}. The particles are collected in 
each class take into account the fractal spectrum

\begin{eqnarray}
&&h-1=1-\nu,\;\;\;\; 0 < \nu < 1;\;\;\;\;\;\;\;\;
 h-1=\nu-1,\;
\;\;\;\;\;\; 1 <\nu < 2;\nonumber\\
&&h-1=3-\nu,\;\;\;\; 2 < \nu < 3;\;\;\;\;\;\;\;\;
 h-1=\nu-3,\;
\;\;\;\;\;\; 3 <\nu < 4;etc.
\end{eqnarray}

\noindent and the spin-statistics relation $\nu=2s$, 
valid for such fractons. The fractal spectrum establishes a 
connection between $h$ and the spin $s$ of the particles: 
$h=2-2s$, $0\leq s\leq \frac{1}{2}$. Thus, there exists a mirror 
symmetry behind this notion of fractal spectrum. 
Given the statistical weight for these classes of fractons

\begin{equation}
\label{e11}
{\cal W}[h,n]=\frac{\left[G+(nG-1)(h-1)\right]!}{[nG]!
\left[G+(nG-1)(h-1)-nG\right]!}
\end{equation}

and from the condition of the entropy be a maximum, we obtain 
the fractal distribution function\cite{R2}

\begin{eqnarray}
\label{e.44} 
n[h]=\frac{1}{{\cal{Y}}[\xi]-h}
\end{eqnarray}

The function ${\cal{Y}}[\xi]$ satisfies the equation 

\begin{eqnarray}
\label{e.4} 
\xi=\biggl\{{\cal{Y}}[\xi]-1\biggr\}^{h-1}
\biggl\{{\cal{Y}}[\xi]-2\biggr\}^{2-h},
\end{eqnarray}

\noindent with $\xi=\exp\left\{(\epsilon-\mu)/KT\right\}$. We understand the 
fractal distribution function as a quantum-geometrical 
description of the statistical laws of nature, 
since the quantum path is a fractal curve and this 
reflects the Heisenberg uncertainty principle.

We can obtain for any class its distribution function considering 
Eq.(\ref{e.44}) and Eq.(\ref{e.4}). For example, 
the universal class $h=\frac{3}{2}$ with distinct values of spin 
$\biggl\{\frac{1}{4},\frac{3}{4},\frac{5}{4},\cdots\biggr\}_{h=\frac{3}{2}}$, 
has a specific Fractal distribution

\begin{eqnarray}
n\left[\frac{3}{2}\right]=\frac{1}{\sqrt{\frac{1}{4}+\xi^2}}.
\end{eqnarray}

\noindent We also have
 
\begin{eqnarray}
\xi^{-1}=\biggl\{\Theta[{\cal{Y}}]\biggr\}^{h-2}-
\biggl\{\Theta[{\cal{Y}}]\biggr\}^{h-1}
\end{eqnarray}

\noindent where

\begin{eqnarray}
\Theta[{\cal{Y}}]=
\frac{{\cal{Y}}[\xi]-2}{{\cal{Y}}[\xi]-1}
\end{eqnarray}

\noindent is the single-particle partition function. 
We verify that the classes $h$ satisfy a duality symmetry defined by 
${\tilde{h}}=3-h$. So, fermions and bosons come as dual particles. 
As a consequence, we extract a fractal 
supersymmetry which defines pairs of particles $\left(s,s+\frac{1}{2}\right)$. 
In this way, the fractal distribution function appears as 
a natural generalization of the fermionic and bosonic 
distributions for particles with braiding properties. Therefore, 
our approach is a unified formulation 
in terms of the statistics which each universal class of 
particles satisfies: from a unique expression 
we can take out any distribution function. In some sense , we can say that 
fermions are fractons of the class $h=1$ and 
bosons are fractons of the class $h=2$.

The free energy for particles in a given quantum state is expressed as

\begin{eqnarray}
{\cal{F}}[h]=KT\ln\Theta[{\cal{Y}}].
\end{eqnarray}

\noindent Hence, we find the average occupation number

\begin{eqnarray}
\label{e.h} 
n[h]&=&\xi\frac{\partial}{\partial{\xi}}\ln\Theta[{\cal{Y}}].
\end{eqnarray}

\noindent The fractal von Neumann entropy per state in terms of the 
average occupation number is given as\cite{R1,R2} 

\begin{eqnarray}
\label{e5}
{\cal{S}}_{G}[h,n]&=& K\left[\left[1+(h-1)n\right]\ln\left\{\frac{1+(h-1)n}{n}\right\}
-\left[1+(h-2)n\right]\ln\left\{\frac{1+(h-2)n}{n}\right\}\right]
\end{eqnarray}

\noindent and it is associated with the fractal distribution function
(Eq.\ref{e.44}).

The entropies for fermions $\biggl\{\frac{1}{2},
\frac{3}{2},\frac{5}{2},\cdots\biggr\}_{h=1}$
 and bosons $\biggl\{0,1,2,\cdots\biggr\}_{h=2}$, can be recovered promptly
 
\begin{eqnarray}
{\cal{S}}_{G}[1]=-K\biggl\{n\ln n +(1-n)\ln (1-n)\biggr\} 
\end{eqnarray}

\noindent and
 
\begin{eqnarray}
{\cal{S}}_{G}[2]=K\biggl\{(1+n)\ln (1+n)-n\ln n\biggr\}. 
\end{eqnarray}

\noindent Now, as we can check, each universal class $h$ of particles, 
within the interval of definition has its entropy defined 
by the Eq.(\ref{e5}). Thus, for fractons of the self-dual class
$\biggl\{\frac{1}{4},
\frac{3}{4},\frac{5}{4},\cdots\biggr\}_{h=\frac{3}{2}}$, we have
  
\begin{eqnarray}
{\cal{S}}_{G}\left[\frac{3}{2}\right]=K\left\{(2+n)\ln\sqrt{\frac{2+n}{2n}}
-(2-n)\ln\sqrt{\frac{2-n}{2n}}\right\} 
\end{eqnarray}

\noindent and for two more examples, the dual classes 
$\biggl\{\frac{1}{3},
\frac{2}{3},\frac{4}{3},\cdots\biggr\}_{h=\frac{4}{3}}$ and 
$\biggl\{\frac{1}{6},\frac{5}{6},\frac{7}{6},\cdots\biggr\}_{h=\frac{5}{3}}$,

the entropies read as  

\begin{eqnarray}
{\cal{S}}_{G}\left[\frac{4}{3}\right]=K\left\{(3+n)\ln\sqrt[3]{\frac{3+n}{3n}}
-(3-2n)\ln\sqrt[3]{\frac{3-2n}{3n}}\right\} 
\end{eqnarray}

\noindent and

\begin{eqnarray}
{\cal{S}}_{G}\left[\frac{5}{3}\right]=K\left\{(3+2n)\ln\sqrt[3]{\frac{3+2n}{3n}}
-(3-n)\ln\sqrt[3]{\frac{3-n}{3n}}\right\}. 
\end{eqnarray}

\noindent

We have also introduced the topological concept of fractal index, 
which is associated with each class. As we saw, $h$ is a geometrical parameter 
related to the quantum paths of the particles and so, we define\cite{R3} 

\begin{equation}
\label{e.1}
i_{f}[h]=\frac{6}{\pi^2}\int_{\infty(T=0)}^{1(T=\infty)}
\frac{d\xi}{\xi}\ln\left\{\Theta[\cal{Y}(\xi)]\right\}.
\end{equation}

\noindent We obtain for the bosonic class $i_{f}[2]=1$, 
for the fermionic class $i_{f}[1]=0.5$ 
and for some classes of fractons, we have 
$i_{f}[\frac{3}{2}]=0.6$, $i_{f}[\frac{4}{3}]=0.56$, $i_{f}[\frac{5}{3}]=0.656$. 
For the interval of the definition $ 1$$\;$$ \leq $$\;$$h$$\;$$ \leq $$\;$$ 2$, there 
exists the correspondence $0.5$$\;$$ 
\leq $$\;$$i_{f}[h]$$\;$$ \leq $$\;$$ 1$, which signalizes 
the connection between fractons and quasiparticles of the conformal field theories, 
in accordance with the unitary $c$$\;$$ <$$\;$$ 1$ 
representations of the central charge. For $\nu$ even it is defined by 

\begin{eqnarray}
\label{e.11}
c[\nu]=i_{f}[h,\nu]-i_{f}\left[h,\frac{1}{\nu}\right]
\end{eqnarray}

\noindent and for $\nu$ odd it is defined by 

\begin{eqnarray}
\label{e.12}
c[\nu]=2\times i_{f}[h,\nu]-i_{f}\left[h,\frac{1}{\nu}\right],
\end{eqnarray}

\noindent where $i_{f}[h,\nu]$ means the fractal 
index of the universal class $h$ which contains the 
statistical parameter $\nu=2s$ or the particles 
with distinct values of spin, which obey specific 
fractal distribution function. For example, we obtain the results

\begin{eqnarray}
&&c[0]=i_{f}[2,0]-i_{f}[h,\infty]=1;\nonumber\\
&&c[1]=2\times i_{f}[1,1]-i_{f}[1,1]=0.5;etc.
\end{eqnarray}

\noindent We have noted in\cite{R3}, for the first time, 
an unsuspected connection betwenn fractal geometry and conformal field theories, 
which second our considerations is expressed by 
Eqs.(\ref{e.1},\ref{e.11},\ref{e.12}).

In another way, the central charge $c[\nu]$ can be obtained using the 
Rogers dilogarithm function, i.e. 

\begin{equation}
\label{e.16}
c[\nu]=\frac{L[x^{\nu}]}{L[1]},
\end{equation}

\noindent with $x^{\nu}=1-x$,$\;$ $\nu=0,1,2,3,etc.$ and 

\begin{equation}
L[x]=-\frac{1}{2}\int_{0}^{x}\left\{\frac{\ln(1-y)}{y}
+\frac{\ln y}{1-y}\right\}dy,\; 0 < x < 1.
\end{equation}

\noindent Thus, we have established a connection between fractal geometry and 
number theory, given that the dilogarithm function appears 
in this context, besides another branches of mathematics\cite{R8}.

\section{Fractional quantum Hall effect}

Such ideas can be applied in the context of the FQHE. This phenomenon is 
characterized  by the filling factor parameter $f$, and for 
each value of $f$ we have the 
quantization of Hall resistance and a superconducting state 
along the longitudinal direction of a planar system of electrons, which are
manifested by semiconductor doped materials, i.e. heterojunctions, 
under intense perpendicular magnetic fields and lower 
temperatures\cite{R9}.

The parameter $f$ is defined by $f=N\frac{\phi_{0}}{\phi}$, where 
$N$ is the electron number, 
$\phi_{0}$ is the quantum unit of flux and
$\phi$ is the flux of the external magnetic field throughout the sample. 
The spin-statistics relation is given by 
$\nu=2s=2\frac{\phi\prime}{\phi_{0}}$, where 
$\phi\prime$  is the flux associated with the charge-flux 
system which defines the fracton $(h,\nu)$. According to our approach 
there is a correspondence between $f$ and $\nu$, numerically $f=\nu$. 
This way, we verify that the filling factors observed 
experimentally appear into the classes $h$ and from the definition of duality 
between the equivalence classes, we note that the FQHE occurs in pairs 
 of these dual topological quantum numbers\\

 $(f,\tilde{f})=\left(\frac{1}{3},\frac{2}{3}\right), 
\left(\frac{5}{3},\frac{4}{3}\right), \left(\frac{1}{5},\frac{4}{5}\right), 
\left(\frac{2}{7},\frac{5}{7}\right),\left(\frac{2}{9},\frac{7}{9}\right), 
\left(\frac{2}{5},\frac{3}{5}\right), \left(\frac{3}{7},\frac{4}{7}\right), 
\left(\frac{4}{9},\frac{5}{9}\right) etc$.\\

All the experimental data satisfy this symmetry principle. 
 In this way, our formulatiom can 
predicting FQHE, that is, consider the duality 
symmetry discovered by us\cite{R2}. Thus, each Hall state 
is described by a system of 
quasiparticles(fractons) such that for a given value of 
filling factor $f$, 
the spin of the objects which constitute the physical 
system is $s=f/2$. We 
understand here fractons as modelling collective 
excitations of a two-dimensional 
electron gas under special conditions like FQHE.

We also observe that 
our approach, in terms of equivalence 
classes for the filling factors, embodies 
the structure of the modular group as discussed in the literature
\cite{R2,R10}. We have 
that the transitions allowed are 
those generated by the condition $\mid p_{2}q_{1}
-p_{1}q_{2}\mid=1$, 
with $h_{1}=\frac{p_{1}}{q_{1}}$ and $h_{2}=
\frac{p_{2}}{q_{2}}$. For example, we have 
the transitions between the classes

\[
\biggl\{\frac{1}{3},\frac{5}{3},\frac{7}{3},\cdots\biggr\}_{h=\frac{5}{3}};
\biggl\{\frac{2}{5},\frac{8}{5},\frac{12}{5},\cdots\biggr\}_{h=\frac{8}{5}};
\biggl\{\frac{3}{7},\frac{11}{7},\frac{17}{7},\cdots\biggr\}_{h=\frac{11}{7}};\]
\[
\biggl\{\frac{4}{9},\frac{14}{9},\frac{22}{9},\cdots\biggr\}_{h=\frac{14}{9}};
\biggl\{\frac{5}{11},\frac{17}{11},\frac{27}{11},\cdots\biggr\}_{h=\frac{17}{11}};
\biggl\{\frac{6}{13},\frac{20}{13},\frac{32}{13},\cdots\biggr\}_{h=\frac{20}{13}} etc.
\]

\noindent This way, we define the universality classes of the quantum Hall transitions, 
which are labeled by the fractal parameter $h$. The topological character 
of these quantum numbers comes from the relation between $h$ and $f$, by the fractal spectrum.

\section{Number theory}

We observe again that our formulation to 
the universal class $h$ of particles with any values of spin $s$ 
establishes a connection between Hausdorff dimension $h$ and 
the central charge $c[\nu]$. Besides this, we have obtained a relation between the 
fractal parameter and the Rogers dilogarithm function, through the 
concept of fractal index, which is defined 
in terms of the partition function associated with each universal class of particles. 
As a result, we have a connection between fractal geometry and number theory. 
Thus,

\begin{eqnarray}
c[\nu]&=&\frac{L[x^{\nu}]}{L[1]}=
i_{f}[h,\nu]-i_{f}\left[h,\frac{1}{\nu}\right],\; 
\nu=0,2,4,etc.\\
c[\nu]&=&\frac{L[x^{\nu}]}{L[1]}=
2\times i_{f}[h,\nu]-i_{f}\left[h,\frac{1}{\nu}\right],\;
\nu=1,3,5,etc.
\end{eqnarray}

\noindent Also we have established a connection between the fractal 
parameter $h$ and the Farey 
sequences of rational numbers. Now, the fractal curve is continuous 
and nowhere differentiable, it is self-similar, it does not depend 
on the scale and has fractal dimension just in the interval 
$1$$\;$$ < $$\;$$h$$\;$$ <$$\;$$ 2$. Given a closed path 
with length $L$ and resolution $R$, the fractal properties of this curve 
can be determined by $h-1=\lim_{R\rightarrow 0}\frac{\ln{L/l}}{\ln R}$, 
where $l$ is the usual length for the resolution $R$ and 
the curve is covering with $l/R$ spheres of diameter $R$.

Farey series $F_{n}$ of order $n$ is the increasing sequence of 
irreducible fractions in the range $0-1$ whose 
denominators do not exceed $n$. They satisfy the properties

P1. If $h_{1}=\frac{p_{1}}{q_{1}}$ and 
$h_{2}=\frac{p_{2}}{q_{2}}$ are two consecutive fractions 
$\frac{p_{1}}{q_{1}}$$ >$$ \frac{p_{2}}{q_{2}}$, then 
$|p_{2}q_{1}-q_{2}p_{1}|=1$.

P2. If $\frac{p_{1}}{q_{1}}$, $\frac{p_{2}}{q_{2}}$,
$\frac{p_{3}}{q_{3}}$ are three consecutive fractions 
$\frac{p_{1}}{q_{1}}$$ >$$ \frac{p_{2}}{q_{2}} 
$$>$$ \frac{p_{3}}{q_{3}}$, then 
$\frac{p_{2}}{q_{2}}=\frac{p_{1}+p_{3}}{q_{1}+q_{3}}$.

P3. If $\frac{p_{1}}{q_{1}}$ and $\frac{p_{2}}{q_{2}}$ are 
consecutive fractions in the same sequence, then among 
all fractions\\
 between the two, 
$\frac{p_{1}+p_{2}}{q_{1}+q_{2}}$
 is the unique reduced
fraction with the smallest denominator.

We have the following 

{\bf Theorem}\cite{R6}: {\it The elements of the Farey series $F_{n}$ 
of the order $n$, belong to the fractal sets, whose Hausdorff 
dimensions are the second fractions of the fractal sets. The 
Hausdorff dimension has values within the interval 
$1$$\;$$ < $$\;$$h$$\;$$ <$$\;$$ 2$, which are associated with fractal curves.}

We observe that the sets obtained are dual sets and, in particular, we have 
a fractal selfdual set, with Hausdorff dimension $h=\frac{3}{2}$. 
In this way, taking into account the 
fractal spectrum and the duality symmetry between sets,  
we can extract for any Farey series of rational numbers, 
fractal sets whose Hausdorff dimension is the second fraction of the set.

\section{Conclusions}

We have introduced a unified description of particles with distinct 
values of spin in terms of their statistics. From a unique expression, 
the fractal distribution function, we can take out distribution functions 
for any universal class $h$ of particles. The Hausdorff dimension of 
the fractal quantum paths of the fractons 
are determined by the fractal spectrum. We have here a 
quantum-geometrical description of the statistical laws of nature.

We verify along these ideas that the FQHE occurs in pairs of dual filling factors. 
These quantum numbers get their topological 
character from the Hausdorff dimension, a geometrical parameter 
associated with the fractal curves of the particles. 
We can check that all experimental results satisfy the symmetry 
principle discovered by us, the duality symmetry betwenn 
universal classes $h$ of particles. The idea of supersymmetry, in some sense, 
appears in this context of the condensed matter and the universality classes 
of the quantum Hall transitions are established.

We emphasize that our formulation is supported by symmetry principles: mirror symmetry 
behind the fractal spectrum, duality symmetry betwenn classes $h$ of particles, 
fractal supersymmetry, modular group behind the quantum Hall transitions.

In another direction, we have established a connection 
betwenn Number Theory and Physics relating fractal geometry and dilogarithm function 
through the concept of fractal index. Also we have determined an algorithm 
for computation of the Hausdorff dimension of any fractal set related to 
the Farey sequences of rational numbers.

Finally, we are thinking about fracton quantum computing 
from the possible perspective of fractons qubits.


\begin{thebibliography}{99}
\bibitem{R1} W. da Cruz, Physica {\bf A313} (2002), 446.
\bibitem{R2} W. da Cruz, Int. J. Mod. Phys. {\bf A15} (2000), 3805.
\bibitem{R3} W. da Cruz and R. de Oliveira, Mod. 
Phys. Lett. {\bf A15} (2000), 1931.
\bibitem{R4} W. da Cruz, J. Phys: Cond. Matter. {\bf 12} (2000), L673.
\bibitem{R5} W. da Cruz, Mod. 
Phys. Lett. {\bf A14} (1999), 1933.
\bibitem{R6} W. da Cruz, Chaos, 
Solitons and Fractals {\bf 17} (2003), 975;\\
W. da Cruz, cond-mat/0301587.
\bibitem{R7} A. M. Polyakov, in {\it Proc. Les
 Houches Summer School
 {\bf vol. IL}}, ed. E. Br\'ezin and J. Zinn-Justin
  (North Holland, 1990) 305.
\bibitem{R8} A. Kirillov, Prog. Theor. Phys. Suppl. {\bf 118} (1995), 61.
\bibitem{R9} R. B. Laughlin, Rev. Mod. Phys. {\bf 71}, (1999), 863;\\
H. Stormer, Rev. Mod. Phys. {\bf 71}, (1999), 875;\\
D. C. Tsui, Rev. Mod. Phys. {\bf 71}, (1999), 891;\\ 
and references therein.
\bibitem{R10} B. P. Dolan, J. Phys. {\bf A32} (1999), L243. 
Nucl. Phys. {\bf B554} (1999), 487.
\end{thebibliography}
\end{document}